\documentclass[twocolumn,pre]{revtex4}
\usepackage{epsfig,graphicx}

\begin{document}

\title{Orientational transitions in symmetric diblock copolymers on
rough surfaces}

\author{Yoav Tsori}
\affiliation{Department of Chemical Engineering, Ben Gurion
University, Beer Sheva 84105, Israel, {\textrm and}\\ UMR 168
CNRS/Institut Curie, 11 rue Pierre et Marie Curie, 75231 Paris CEDEX
05, France. E-mail: tsori@bgu.ac.il}
\author{Easan Sivaniah}
\affiliation{ Department of Chemical Engineering, Texas Technical
University, Lubbock, TX 79409-3121 USA. E-mail:
easan.sivaniah@coe.ttu.edu}
\author{David Andelman$^*$}
\affiliation{ School of Physics and Astronomy, Raymond and Beverly
Sackler Faculty of Exact Sciences, Tel Aviv University, Tel Aviv
69978, Israel. E-mail: andelman@post.tau.ac.il}
\author{Takeji Hashimoto}
\affiliation{Department of Polymer Chemistry, Graduate School of
Engineering, Kyoto University, Katsura, Kyoto 615-8510, Japan.
E-mail: takeji@alloy.polym.kyoto-u.ac.jp}
\date{15/4/2005 - submitted to Macromolecules}

\begin{abstract}
We present a model addressing the orientation transition of
symmetric block copolymers such as PS/PMMA on smooth and rough
surfaces. The distortion free energy of parallel and perpendicular
lamellar phases in contact with a rough solid surface is calculated
as function of the surface roughness amplitude and wavelength, as
well as the polymer lamellar periodicity (molecular weight). We find
an analytical expression for the orientation transition. This
expression is compared and agrees well with recent experiments done
with six different  polymer molecular weights and surface
preparations.
\end{abstract}

\maketitle

As self-assembling systems become better understood, more emphasis
is being given to finding ways to control the assembled structure,
i.e. to orient ordered phases in a certain direction or anneal
defects \cite{thomas_rev,grier,sancaktar}. Block copolymers (BCP)
are excellent model systems, which provide a good balance between
price and chemical versatility, and are being extensively studied
for technological applications as well as from a basic scientific
viewpoint \cite{leibler,schick}. There are numerous ways to affect
the BCP phase-behavior and orientation. For example, using shear
flow \cite{GHF}, confinement between two solid surfaces
\cite{muthu1,TA1}, or application of an external electric field
\cite{AH93,russell,krausch,muthu2,TA2}.

In this note we consider a lamellar phase of symmetric diblock
copolymers (each block has a mole fraction of $f=0.5$) on top of a
rough surface. The amplitude and periodicity of surface modulations
determine if the lamellae will be parallel or perpendicular to the
substrate \cite{TJ,chakra}, as has been recently shown
experimentally by Sivaniah et al \cite{hash1,hash2}. This new and
alternative method to orient BCPs can be advantageous to the methods
mentioned above because of its simple experimental setup. The aim of
this Note is to extend results of a previous theoretical modelling
\cite{TA3} showing its direct applicability to the experimental
findings \cite{hash1,hash2}.

The surface roughness is modelled by a single one-dimensional
corrugation mode, whose height in the $z$-direction above an $(x,y)$
reference plane is given by $h(x)=R\cos(q_sx)$. As is shown on
Fig.~1, $q_s$ and $R$ are the wavenumber and amplitude of the
surface roughness, respectively. The BCP is put above the substrate
in the half-space $z\geq h(x)$. In addition, $q_0=2\pi/D$ is the
wavenumber of the bulk lamellae of width $D$, $\gamma_{\rm AB}$ is
the interfacial interaction (per unit area) between the A and B
blocks in the polymer chain, and $\delta=\gamma_{_{\rm
subs,A}}-\gamma_{_{\rm subs,B}}$ is the surface tension difference
between the substrate and the two types of polymer blocks.

\begin{figure}[h!]
\begin{center}
\includegraphics[scale=0.50,bb=70 370 510 580,clip]{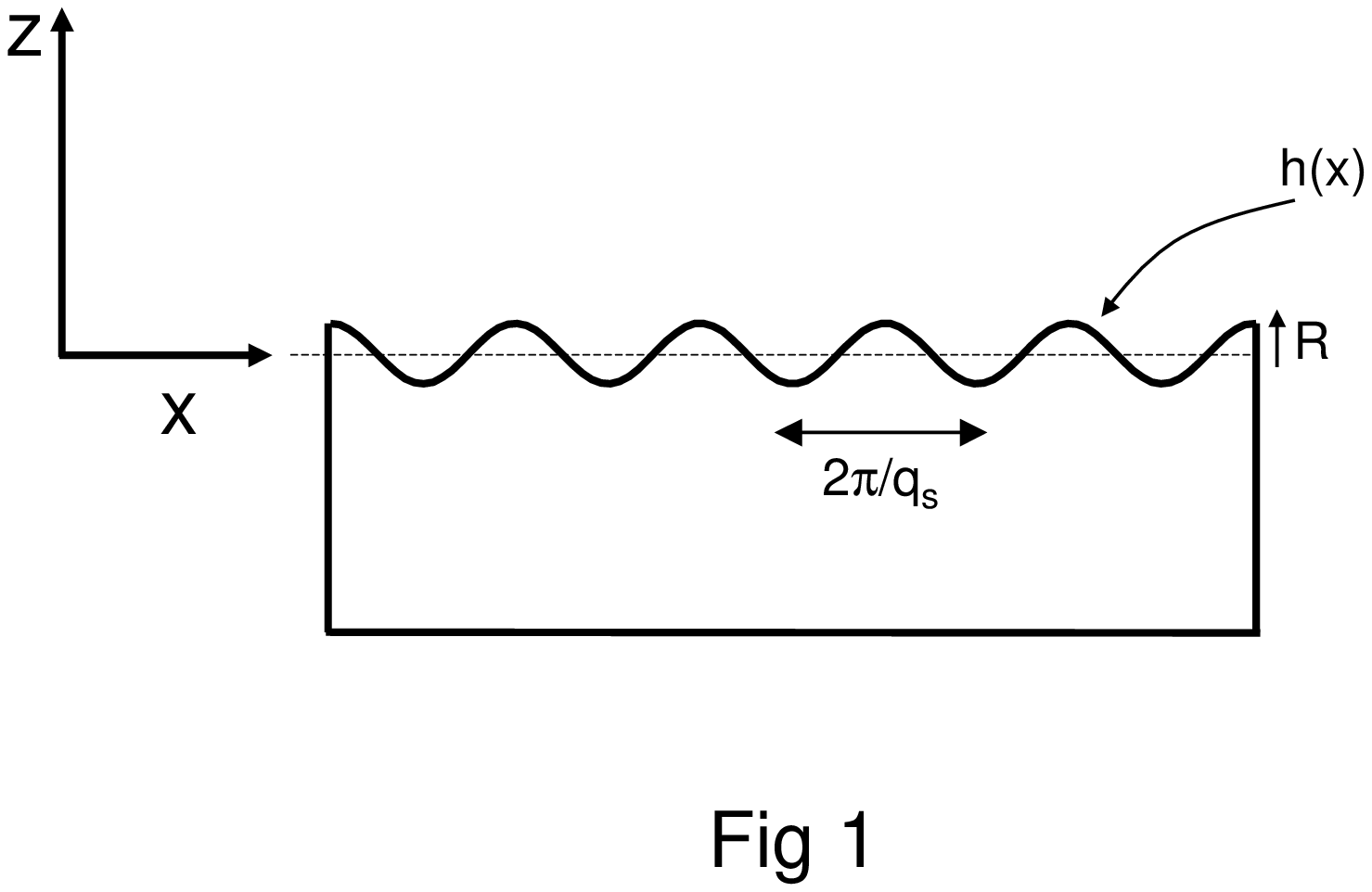}
\end{center}
\caption{\textsf{Schematic illustration of the rough confining
surface.}}
\end{figure}

We start by examining the order-parameter of lamellae oriented
perpendicular to the surface. The presentation follows the same
lines as of Ref.~\cite{TA3}:
\begin{eqnarray}\label{phi_perp}
\phi_\perp({\bf r})=\phi_{0}\cos\left(q_0x+q_0u(x,z)\right)
\end{eqnarray}
This is the deviation of the A-monomer relative concentration from
its average value $f=0.5$. The amplitude of sinusoidal variation,
$\phi_0$, depends on the degree of segregation and vanishes at the
Order-Disorder Temperature (ODT). The function $u(x,z)$ is a
slowly-varying function that describes surface-induced perturbations
of the lamellae from their perfect shape. We write the bulk part of
the free energy, in complete analogy with the elastic energy of
smectic liquid crystals \cite{TJ,PGGP}:
\begin{eqnarray}
F_b=\frac12\int\left[K\left(u_{zz}\right)^2+B\left(u_x\right)^2\right]
{\rm d}^3r
\end{eqnarray}
where $u_x=\partial u/\partial x$, $u_{zz}=\partial^2u/\partial
z^2$, $K\sim D\gamma_{\rm AB}$ is the bending modulus and
$B\sim\gamma_{\rm AB}/D$ is the compression modulus.

Several assumptions are made regarding the length scales and
energies involved, as are explained in more detail in Ref.
\cite{TA3}:
\begin{equation}\label{valid1}
1> q_0R>q_sR>(q_0R)^{3/2}
\end{equation}
and
\begin{equation}\label{valid2}
\phi_0\delta\ll\sqrt{BK}\simeq\gamma_{\rm AB}
\end{equation}

We will be mainly interested in the poly(styrene)/
poly(methylmethacrylate) system, where the A-block is chosen as the
PS and the B-block as PMMA. The corresponding parameters are
$\delta\simeq 0.25$ mN/m for the surfaces considered below, and
$\gamma_{\rm AB}=1$ mN/m, so eq~\ref{valid2} roughly holds. The
inequalities in eq~\ref{valid1} are  not satisfied in all the
experiments. While $q_0R$ is indeed larger than $q_sR$, $q_0R$ is
between 1 and 4 and is not smaller than unity as assumed. Therefore,
$q_0R>(q_0R)^{3/2}$ does not strictly hold. Nevertheless, using the
above inequalities we were able to make simple analytical
predictions by minimizing the energy with respect to the distortion
field $u$. Up to numerical prefactors, the bulk free energy of the
perpendicular state is \cite{TA3}
\begin{eqnarray}\label{Fperp_old}
\frac{F_\perp^0}{S}\sim\frac{\delta^2\phi_0^2}{K}\frac{1}{q_0}
\end{eqnarray}
where $S$ is the surface area.

We repeat the same calculation as above but now for parallel
lamellae. The order parameter is given by
\begin{eqnarray}
\phi_{_\parallel}({\bf r})=-\phi_0\cos(q_0z+q_0u(x,z))
\end{eqnarray}
and the bulk free-energy is
\begin{eqnarray}
F_b=\frac12\int\left[K\left(u_{xx}\right)^2+B\left(u_z\right)^2\right]
{\rm d}^3r
\end{eqnarray}
We minimize the distortion field $u$ in the same limits as in
eqs~\ref{valid1} and \ref{valid2}, and find \cite{TA3}
\begin{eqnarray}\label{Fpara_old}
\frac{F_{_\parallel}^0}{S}\sim\frac{\delta^2\phi_0^2}{K}\frac{1}{q_0}\left(\frac{q_0}{q_s}
\right)^2\left(q_0R\right)^2
\end{eqnarray}

Equations~\ref{Fperp_old} and \ref{Fpara_old},  use the surface
energy to obtain the distortion field $u$. We  now add this
substrate-BCP interfacial tension and compare the gain and loss in
the total free energy of the two states, including the bulk
distortion and interfacial tension terms. This was not done in
Ref.~\cite{TA3}. In the case of perpendicular lamellae, the
substrate is approximately equally covered by the A and B monomers
(the symmetric case of PS/PMMA). Hence, adding the
interfacial-tension term to eq~\ref{Fperp_old} results in the
following free-energy
\begin{eqnarray}\label{Fperp}
\frac{F_\perp}{S}&\simeq&\frac{\delta^2\phi_0^2}{K}\frac{1}{q_0}\nonumber\\&+&
\frac12\left(\gamma_{_{\rm subs,A}}+\gamma_{_{\rm subs,B}}\right)
\left(1+\frac14(q_sR)^2\right)
\end{eqnarray}
The extra factor $1+\frac14(q_sR)^2$ is the ratio
between the real surface profile $h(x)=R\cos(q_sx)$ and the flat one,
$h=0$, for small surface corrugations.

For parallel lamellae, we have a surface in contact with  a layer
rich in B monomers (PMMA). Neglecting surface proximity effects, we
consider that this layer has a concentration of B-monomer with
amplitude $\frac12-\phi_0$ and A-monomers (PS) with amplitude
$\frac12+\phi_0$, recalling that $\phi_0$ is the deviation of the
order parameter from $\frac12$. The energy is
$(\frac12-\phi_0)\gamma_{_{\rm
subs,A}}+(\frac12+\phi_0)\gamma_{_{\rm subs,B}}$, and the total
parallel free-energy becomes
\begin{eqnarray}\label{Fpara}
\frac{F_{_\parallel}}{S}\simeq\frac{\delta^2\phi_0^2}{K}\frac{1}{q_0}
\left(\frac{q_0}{q_s}\right)^2\left(q_0R\right)^2+
\left[\left(\frac12-\phi_0\right)\gamma_{_{\rm subs,A}}\right. \nonumber\\
\left.+\left(\frac12+\phi_0\right)\gamma_{_{\rm subs,B}}\right]
\left(1+\frac14(q_sR)^2\right)\nonumber\\
\end{eqnarray}

In order to find the orientation transition, we equate
$F_{_\parallel}$ to $F_\perp$ [eqs~\ref{Fperp} and \ref{Fpara}]
while estimating  $Kq_0\simeq 2\pi\gamma_{\rm AB}$. The transition
value of $(q_sR)^2$ is given by:
\begin{equation}\label{gov_eq}
(q_sR)^2={{\phi_0\delta+2\pi\gamma_{\rm AB}}\over{\phi_0\delta(q_0/
q_s)^4-{\pi\over 2}\gamma_{\rm AB}}}
\end{equation}
Note that this equation includes the information of the melt
segregation via $\phi_0$ ($|\phi_0|<\frac12$). Naturally, as the
temperature approaches the ODT, $\phi_0$ tends to zero the energetic
difference between the parallel and perpendicular states goes to
zero as well.

In order to compare these predictions of the lamellar orientation
dependence on the various roughness parameters, we used some of the
results reported in Ref.~\cite{hash2}. In that paper sample
orientation was determined by a combination of cross-sectional TEM
microscopy, atomic force microscopy and dynamic secondary-ion mass
spectroscopy. The principle result was to demonstrate that an
increase in the substrate roughness amplitude, $R$, led to a
transition from parallel to perpendicular orientation, while other
substrate parameters were untouched.

Six observations of orientation were also made with three molecular
weights of symmetric PS-PMMA block copolymer (of different $q_0$) on
four substrates of different $q_s$. We can use these observations to
test the validity of our current theoretical model. The BCP samples
are denoted 18k-18k, 38k-36.8k and 50k-54k, according to the
molecular weight of PS and PMMA blocks in the chain, respectively.
The four substrates are: super-critically rough Indium Tin Oxide
(SC-ITO), under-critically rough ITO (UC-ITO), smooth ITO (S-ITO)
and super-critically rough polyimide (SC-PIM). All the experimental
parameters are summarized in Table I.  The prefixes of $super-$ and
$under-$ were used to denote the degree of roughness of the
substrates.  The SC-PIM substrate was made by imprinting a polyimide
surface with a SC-ITO surface. Therefore, SC-PIM and SC-ITO had
identical topological features. Contact angle measurements at
$200^{\circ}$C on all of the substrates revealed that there was no
large difference in the wetting properties of PS and PMMA on all of
these substrates. For more details see Ref.~\cite{hash2}.

\begin{figure}[h!]
\begin{center}
\includegraphics[scale=0.73,bb=180 60 430 730,clip]{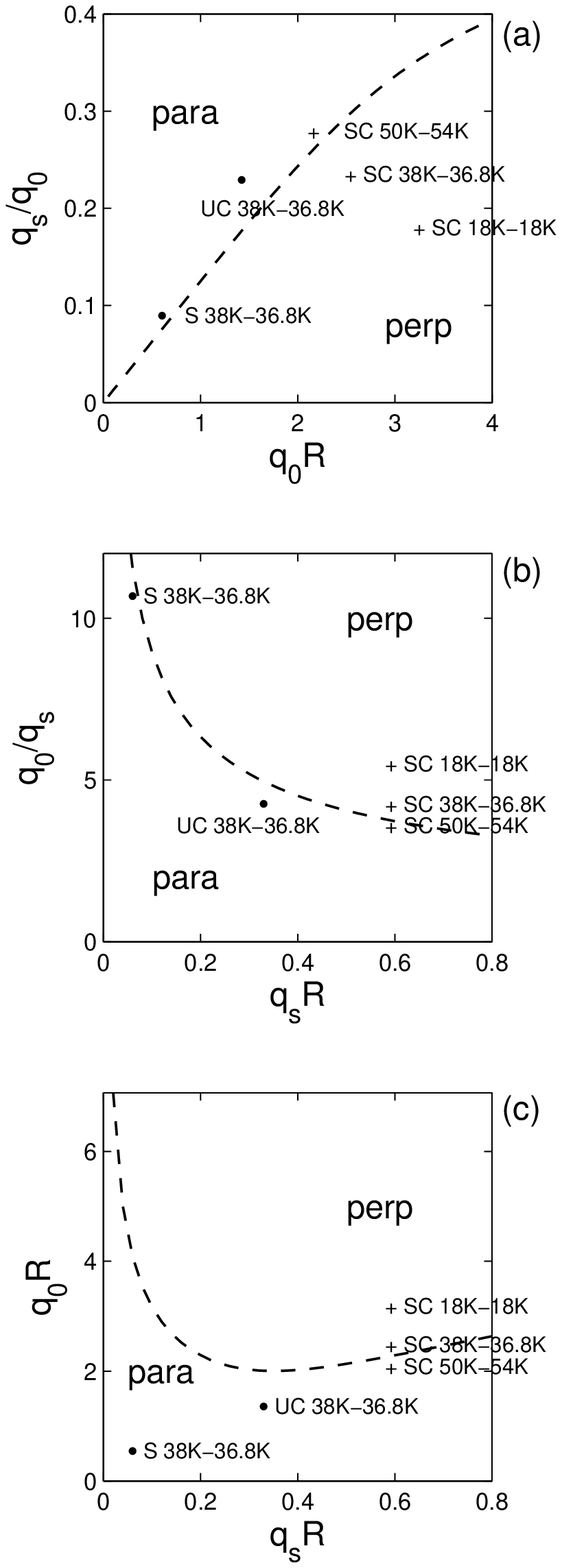}
\end{center}
\caption{\textsf{Comparison between the model and experimental phase
diagram for the perpendicular and parallel lamellar configurations
on a rough substrate. The solid line is calculated from
eq.~\ref{gov_eq}. (a) $R$ and $q_s$ are scaled by $q_0$, the
lamellae wavenumber. The  ``+'' data points correspond to observed
perpendicular morphology, while the ``$\bullet$" ones to parallel
lamellae (see also Table I). (b) A different plot of the same
theoretical prediction from eq.~\ref{gov_eq} and experimental
points. The surface wavenumber $q_s$ is used to scale $q_0$ and $R$.
(c) The surface amplitude $R$ is used to scale $q_0$ and $q_s$. In
all three plots $\gamma_{\rm AB}=1$ mN/m, $\delta=0.25$ mN/M and
$\phi_0=0.4$. }}

\end{figure}

An assumption of identical substrate surface energy allows all six
observations to be collated onto a single orientational phase
diagram. In Fig. 2 we plot the experimental points and the
transition lines predicted by eq~\ref{gov_eq} on three types of
plots. In part (a), $q_0=2\pi/D$ is used to scale the two other
parameters: $q_s$ and $R$ and to produce two dimensionless
parameters for the plot: $q_s/q_0$ and $q_sR$. In the range of
experimental parameters, the transition line between parallel and
perpendicular states, eq~\ref{gov_eq}, is very close to a straight
line (up to about $q_0R\simeq 2.5$ in Fig.~2(a)). Namely, for a
fixed $q_0$, $ q_s\sim R$.

Our above findings  should be compared with a previous model by
Turner and Joanny \cite{TJ} that predicted an orientation transition
at $q_s\sim 1/R$ and independent of $q_0$. Because of the limited
number of experimental systems, it is hard to rule out any
theoretical fit. However, it looks that the latter prediction does
not fit well the experiments summarized in Fig.~2. In our model the
limit of $q_s\sim R$ is obtained by formally setting $q_0\to 0$ in
eq~\ref{gov_eq}. However, we note that the validity of our model, as
well as the experiments, is in the opposite limit of $q_0>q_s$.
Figure~2 (a) shows that all samples found to be in the parallel
morphologies in the experiments lie indeed above the theoretical
transition line and all perpendicular morphologies lie below it,
with the exception of the SC-ITO 50k-54k sample, which lies a little
inside the parallel region, although it is measured as a
perpendicular state.

The same information is presented differently in parts (b) and (c)
of Fig.~2. In (b) we think of $q_s$ as the rescaling factor and plot
$q_0/q_s$ as function of $q_sR$. From eq~\ref{gov_eq} and under the
condition $\gamma_{\rm AB}\ll\phi_0\delta$, we get $q_0/q_s \simeq
(q_sR)^{-1/2}$. In Fig.~2 (b) we see again that beside the SC-ITO
50k-54k sample, all other data points fit with the theoretical
prediction. And finally in Fig.~2 (c), $q_0$ and $q_s$ are rescaled
by $R$.

The phase diagrams in parts (b) and (c) may seem counter-intuitive
at first sight. At a given surface roughness $q_sR$, the transition
from  the parallel phase to the perpendicular one occurs as $D$
decreases (or $q_0$ increases).
This surprising behavior can be understood by looking at the
dependence of the distortion field $u(x,z)$ on the distance $z$ from
the surface. For perpendicular lamellae, $u\sim \exp({-k_\perp z})$,
where $k_\perp\sim1/D$. Hence distortions relax at a distance from
the substrate comparable to the lamellar spacing.
Undulations in the parallel phase, however, are given by $u\sim
\exp({-k_{_\parallel} z})$, where $k_{_\parallel}=q_s^2/q_0$. As the
BCP molecular weight decreases, $D$ decreases, $q_0\sim D^{-1}$
increases, and  $k_{_\parallel}\sim D$ decreases, resulting in a
longer extent of the distortion $u$ field in the $z$ direction.
Hence, the accumulated frustration of parallel lamellae leads to
their relative instability towards the perpendicular phase. In other
words, the smaller the molecular weight is, the more stable the
perpendicular lamellae tend to be.

More experiments should be carried out so that the phase-diagram can
be fully mapped. In particular the above-mentioned effect should be
further explored, giving special attention to the possible creation of
island and holes or other defects which are not included in this
simple theoretical framework.

\section*{Acknowledgements}
Partial support from 21st century COE program, COE for a United
Approach to New Materials, the United States-Israel Binational
Science Foundation (BSF) under grant No. 287/02 and the Israel
Science Foundation under grant No. 210/01 is gratefully
acknowledged.



\begin{widetext}
\begin{table}[h!]
\begin{center}
\begin{tabular}{|c|c|c|c|c|c|}
\hline
& & & & & \\
surface & $q_s$ [nm$^{-1}$] & $R$[nm] & 18k-18k  & 38k-36.8k
 & 50k-54k
\\
& & & $q_0=0.22 {\rm ~nm}^{-1}$ & $q_0=0.17 {\rm ~nm}^{-1}$ &
$q_0=0.14 {\rm ~nm}^{-1}$
\\
& & & $D=28.6$~nm &$D=36.7$~nm & $D=43.5$~nm\\ \hline
\hline
rough SC--ITO & 0.04 &  14.5& perp & perp & perp\\
\hline
rough UC--ITO & 0.04 &  8&  & para & \\
\hline
smooth S--ITO & 0.016 &  3.2&  & para & \\
\hline
rough SC--PIM & 0.04 &  14.5&  & perp & \\
\hline
\end{tabular}
\end{center}
\caption{\textsf{Experimental results from Ref.~\cite{hash2} for
different PS/PMMA samples and different rough surfaces. Left column
indicates the type of substrate used, $q_s$ and $R$ are the
corrugation wavenumber and amplitude, respectively (see also
Fig.~1). The name of a sample indicates the molecular weight of the
PS/PMMA blocks. The morphology is given for the six experiments that
were carried out. }}
\end{table}

\end{widetext}


%

\end{document}